%% file: moon2.tex
\documentclass [12pt,twoside]{article}           

\input cpreb.tex

\def\qvec#1{\overrightarrow{#1}}

\begin{document}
\thispagestyle{empty}



\markboth{{\sl \hfill  \hfill \protect\phantom{3}}}
        {{\protect\phantom{3}\sl \hfill  \hfill}}

\color{yellow} 
\hrule height 40mm depth 1mm width 170mm 
\color{black}
\vskip -33mm
\centerline{\bf \LARGE \color{blue} How can one detect}
\vskip 5mm
\centerline{\bf \LARGE \color{blue} the rotation of the Earth
``around the Moon''? }
\vskip 5mm
\centerline{\bf \LARGE \color{blue} Part 2: Ultra-slow fall}
\vskip 16mm
\centerline{\bf \large 
Bertrand M. Roehner$ ^{1,2} $ }

\large
\vskip 8mm

{\bf Abstract}\quad The paper proposes an alternative
to the Foucault pendulum for detecting various movements of
rotation of the Earth. Calculations suggest that if the
duration of a ``free'' fall becomes longer the
eastward deflection will be amplified in proportion
with  the increased duration. 
Instead of 20 micrometers for a one-meter fall,
one can expect deflections more than 1,000 times larger
when the fall lasts a few minutes.\qL
The method proposed in this paper
consists in using the buoyancy of a 
(non viscous) liquid in order to work in reduced gravity. 
In a liquid of density $ \rho $, the gravity $ g $
is replaced by a virtual gravity
$ g'=g(1-\rho/\rho_1) $ where $ \rho_1 $ is the
density of the falling body. \qL
Not surprisingly, as in many astronomical observations, 
the main challenge is to minimize the
level of ``noise''. Possible sources of noise are discussed
and remedies are proposed.\qL
In principle, the experiment should be done in superfluid
helium. However, a preliminary experiment done in water gave
encouraging results in spite of a fairly high level of noise.
In forthcoming experiments the main objective will be to identify
and eliminate the main sources of noise.\qL
This experiment differs from the Foucault pendulum by
its greater flexibility.
By adequately selecting the major parameters, e.g. duration of the
fall, viscosity of the fluid, size of the falling body,
one can change the deflection target.\qL
It is hoped that the present paper
will encourage new experiments in this direction.
\vskip 3mm

\centerline{\it 11 December 2011.
Preliminary version, comments are welcome}
\vskip 3mm

\vskip 2mm

{\normalsize 
1: Department of Systems Science, Beijing Normal University, Beijing,
China. \qL
2: Email address: roehner@lpthe.jussieu.fr.
On leave of absence from the ``Institute for Theoretical and
High Energy Physics'' of University Pierre and Marie Curie, Paris, France.
}

\vfill \eject

\qI{Introduction}

In this paper we propose an alternative to the Foucault pendulum%
\qfoot{The Foucault pendulum method was discussed in 
the first of the present series of two papers. For convenience
this article will be referred to simply as ``Part 1''.}%
for the detection of movements of rotation of the Earth. 
This method is based on the well-known observation that 
instead of following a vertical trajectory, a falling body
is in fact slightly deflected toward the East. \qL
It can be noted 
that this deflection is the same in the northern and southern 
hemisphere. The reason is because this effect is due to the
{\it horizontal} component of the vector
of angular rotation (see Fig. 2a,b of Part 1), whereas the Foucault
effect is due to the vertical component. The horizontal
component is directed toward the north in both hemispheres.

 \qA{Free fall in the air}
Various experiments performed in the past (Table B1)
show that in air or in vacuum the
eastward deflection is small. 
The theoretical formula discussed
in more detail in the appendix shows that for a fall from a height
of one meter one should expect a deflection of only 22 micrometers.
Clearly, for such a short deflection it is almost impossible to
get a good accuracy. One major problem is the exact determination
of the bottom point which is on the
vertical of the starting point. If one wishes an accuracy
of 1\%, this point must be determined with an accuracy of 0.2 micrometer%
\qfoot{The accuracy is also limited by the dispersion due to
turbulence. However, as explained in Appendix C, this problem
can be overcome (at least in principle) by increasing the number
of falls.}%
.
Consequently, to improve the accuracy of this experiment
one needs to increase the size of the deflection. So the question
becomes: ``How can the eastward deflection be amplified?''

\qA{Free fall in a fluid}
The theoretical formula for the deflection may give us a possible
hint. This formula can be written in different ways. There are
basically 4 variables. 
\qbu The horizontal angular velocity of the earth 
$ \Omega_h = \Omega\cos \lambda $, where $ \lambda $ is the latitude.
\qbu The acceleration of gravity $ g $
\qbu The height of the fall $ h $
\qbu The duration of the fall $ t $
\qpar
The standard free fall formula $ h=(1/2)gt^2 $ allows us
to write the deflection in two alternative forms:
 $$ d= { 2\over 3 }\Omega_ht \qn{1a} $$
 $$ d= { 2\sqrt{2}\over 3 }\Omega_h { h^{3/2}\over g^{1/2} } \qn{1b} $$

Expression (1a) suggest that, for a given height, 
$ d $ can be amplified by increasing 
the duration of the fall. 
\qpar

At first sight this condition may seem surprising.
Indeed, if (for a given height) the fall lasts longer this
means that the velocity is lower, but then the Coriolis
force will also be lower. So, why should one get an
amplified deflection?
\qpar

In fact, the amplification occurs only if the eastward 
movement is uniformly accelerated. This can be seen by  a 
simple argument. We denote by $ v_m $ the
average vertical velocity. Then, 
for the sake of simplicity we replace the vertical 
movement by a stationary movement of velocity $ v_m $.
Under such conditions the 
Coriolis force 
$ \qvec{C}=2m\qvec{v_m}\wedge\qvec{\Omega} $
is constant during the fall
which means that the horizontal movement is
uniformly accelerated%
\qfoot{This is true even for a movement in a fluid, at least
until the drag becomes comparable to the driving force.}
with an acceleration
$ \gamma_c=C/m =Kv_m $.
For such a movement the eastward deflection is:
$$ x={ 1\over 2 }\gamma_c t^2 = 
{ K\over 2 } v_m \left({ h\over v_m }\right)^2 = 
{ K\over 2 }{ h^2\over v_m } $$
 
The last expression shows that the smaller the vertical velocity 
the larger the deflection.
In short, the amplification of the deflection occurs because of the
quadratic time factor. When the vertical velocity is
divided by 2 (for instance), the Coriolis acceleration $ \gamma_c $
is also divided by 2, but thanks to the time factor $ t^2 $ 
the deflection is multiplied by 4.
\qpar

Now we must ask ourselves how a low velocity can be achieved
practically.

\qA{Ultra-slow falls}

How can one obtain a fall with a slow velocity? An answer is
suggested by expression (1b) which shows that the only way is to
reduce $ g $. How can one reduce $ g $?
Apart from doing the experiment in a spacecraft orbiting around
the earth and in which there is a state of microgravity%
\qfoot{ Actually, even apart from its cost, such an experiment may 
not be easy to carry out in an orbiting satellite.
There is no uniform microgravity in a satellite.
In fact, once released an object becomes a satellite in
its own moving in an orbit sightly different from the one
of the satellite itself. The connection between the two orbits
may be fairly complicated, depending  for instance upon 
the respective size of the satellite and of the object.}%
,
an obvious way is to use the buoyancy in order 
to counterbalance $ g $. If a body of density $ \rho_1 $ falls
in a fluid of density $ \rho $, it will experience a
virtual gravity $ g' $ given by the difference between its weight
and the buoyancy: $ g'=g(1-\rho/\rho_1) $. 
If $ \rho_1 $ is adjusted to become almost equal to $ \rho $ 
(but still somewhat larger) $ g' $ can (at least in principle)
be made arbitrarily small. As a result, $ d $ should become
fairly large.

\qA{How can one reduce the effect of friction?}

An objection arises immediately. Indeed, the expressions (1a,b)
are derived from equations which do not take friction into account. 
How can one create buoyancy without
friction? There are (at least) two possible answers.
\qee{1} One way is to make the experiment in a fluid that
has zero viscosity, for instance in superfluid helium-3 at a 
temperature under 2.1 degree Kelvin%
\qfoot{Under standard pressure, helium-3 becomes liquid at
4.2 degree but the property of superfluidity appears only
under 2.1 degree.}%
.
\qee{2} Another way to make friction small in comparison with
the other forces is to increase appropriately the dimensions of
the falling body.
The argument goes as follows.
The friction force is basically proportional to the
area $ A $ of the section of the object that is perpendicular to
the velocity, whereas the Coriolis force 
$ \qvec{C}=2m\qvec{v}\wedge\qvec{\Omega} $, and the
reduced gravity force $ F=m\qvec{g'} $ are both proportional
to the mass of the falling body. Thus, when the body becomes larger
the relative importance of  the friction force is reduced.
\qpar

A confirmation of this argument can be found in the expression
of the asymptotic velocity of a falling body, namely
(Wikipedia 2011a)
$ v_a=\sqrt{2mg/\rho A C_d} $ where $ C_d $ is the drag coefficient%
\qfoot{In a general way
the drag coefficient depends upon the Reynolds
number $ Re $ but it is almost constant 
(and equal to about 0.5) in the range $ 100 < Re < 10^5 $.}%
.
At this velocity
the friction force is equal to the weight minus the buoyancy.
This expression shows that when $ m/A $ becomes large
the asymptotic velocity goes to
infinity as is the case for a fall in vacuum. In other
words, the fall of a big object is almost identical
to what it would be in vacuum. \qL
The 
increase of the ratio $ m/A $ can occur either through an increase
in density or through an increase in volume.
In our case the density is almost fixed because it must
be close to the density of the fluid. Therefore it is
the ratio $ V/A $. At this point one should
observe that we need only to focus on the {\it horizontal} movement%
\qfoot{The movement along the horizontal $ x $ axis is 
described by the equation: 
$$ { md^2x \over dt^2 } = 2m\Omega_h v_z -F_d $$ 

For the sake of simplicity the vertical velocity $ v_z(t) $ can 
be replaced by its average value $ v_m $. In this
approximation the horizontal movement is exactly identical to 
the fall of a body in a fluid.}%
. 
The only role of the vertical movement is to generate the vertical
velocity which is necessary for the Coriolis force to appear.
For the horizontal movement, the ratio $ V/A $ is 
proportional to the length in the horizontal direction.\qL
For instance, if the falling body
is a cylinder it should be a fairly flat one.
We will discuss later how this condition can be implemented in
practice.

\qI{What deflection can one expect?}

The previous argument tells us that the falling module
should be ``rather big'' but it does not say how big it should 
be. Moreover, before starting  experiments one would like to
know what is the magnitude of the deflection that can be expected.
To this aim we made a computer simulation whose results
are summarized in Fig. 1.
%
\begin{figure}[htb]
 \centerline{\psfig{width=12cm,figure=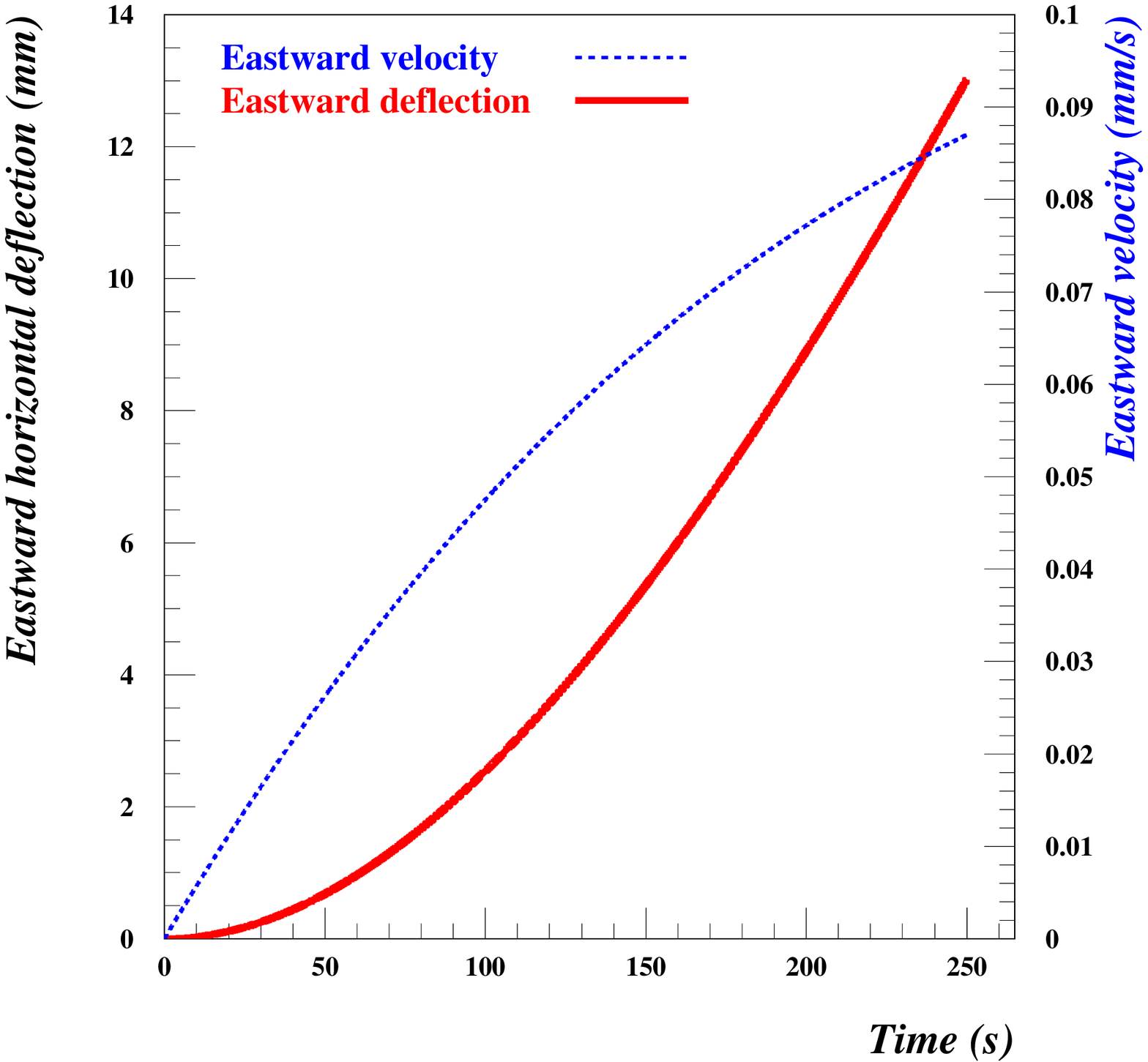}}
\vskip 3mm
\qleg{Fig. 1: Eastward deflection during an ultra-slow fall.}
{The fall occurs in water over a height of 1 meter and has a duration
of 250 seconds. The mass of the falling module is 280 grams.
The velocity curve (broken line) shows that the 
horizontal velocity increases
linearly during a first phase which lasts about one half of the
total time. During this phase the Coriolis force is notably larger
than the drag which results in a fairly constant acceleration.
During this first phase the drag increases along
with the velocity; this leads to a second phase in which the
drag becomes almost equal to the Coriolis force with the result
that the acceleration tends toward zero. During this
second phase the velocity increase becomes slower and slower
as the velocity converges toward its limiting value.}
{}
\end{figure}
%
Before considering these results one needs to realize that
such a simulation can only provide a fairly rough 
picture because the description of the drag is known to be
fairly inappropriate in several respects%
\qfoot{As mentioned in Appendix A, this description
does {\it not} include the effects of turbulence. Moreover,
the drag coefficient is only known in an approximate way.}%
.
In spite of these limitations one can draw the following
conclusions.
\qee{1} The magnitude of the expected deflection is of the
order of several millimeters which is about 100 to 1,000 times larger
than for a similar fall in air.
\qee{2} For a light falling body  the maximum 
of the horizontal velocity
is reached much faster than for a more heavy one.
For a weight of about 1 gram calculations
show that the maximum velocity is reached in 
less than 10 seconds
whereas for a weight of 250 grams the uniformly accelerated
regime lasts  about 250 s. Thus, for a light body it is useless
to use a long falling time because the deflection will not be larger
than for a much shorter falling time.
This comes from the fact that, as observed at the beginning,
unless the movement is uniformly accelerated one does 
not gain anything by increasing the the duration of the fall%
\qfoot{Of course, the deflection continues to increase even
when the velocity has become stationary. But one does not gain
anything in the sense that a
shorter falling time would 
give a higher average vertical velocity and, as a result,
the same deflection would be reached more quickly.}%
.
\qL
The result about the length
of the uniformly accelerated regime can also be seen analytically.
The analytic form of the solution is (Wikipedia 2011\ a)
$$ v_x(t)=\left[{ 2\gamma_c \over \rho C_d }{ m\over A }\right]^{1/2}
\tanh\left(t\sqrt{ { \gamma_c \rho C_d \over 2 }{ A\over m }}\right) $$

As for $ t=1 $ the function $ \tanh t $ is already within 20\%
of its asymptotic value we see that the length of the time interval
for which $ \tanh(at) $ increases linearly is about $ 1/a $.
In other words, the duration of the ``useful'' regime is 
$ \sqrt{(2/\gamma_c \rho C_d)(m/A)} $. We see that it increases 
along with the ratio $ m/A $.

\qI{Implementation of the experiment}

Here, as in many physics experiments, the main challenge is
to reduce the level of noise. What are the main sources of 
noise? 

\qA{Sources of noise}

The following list proceeds
in chronological order from the moment when the fall
starts to the moment when the module
hits the bottom of the tank (Fig. 2a, 2b).
%
\begin{figure}[htb]
 \centerline{\psfig{width=7cm,figure=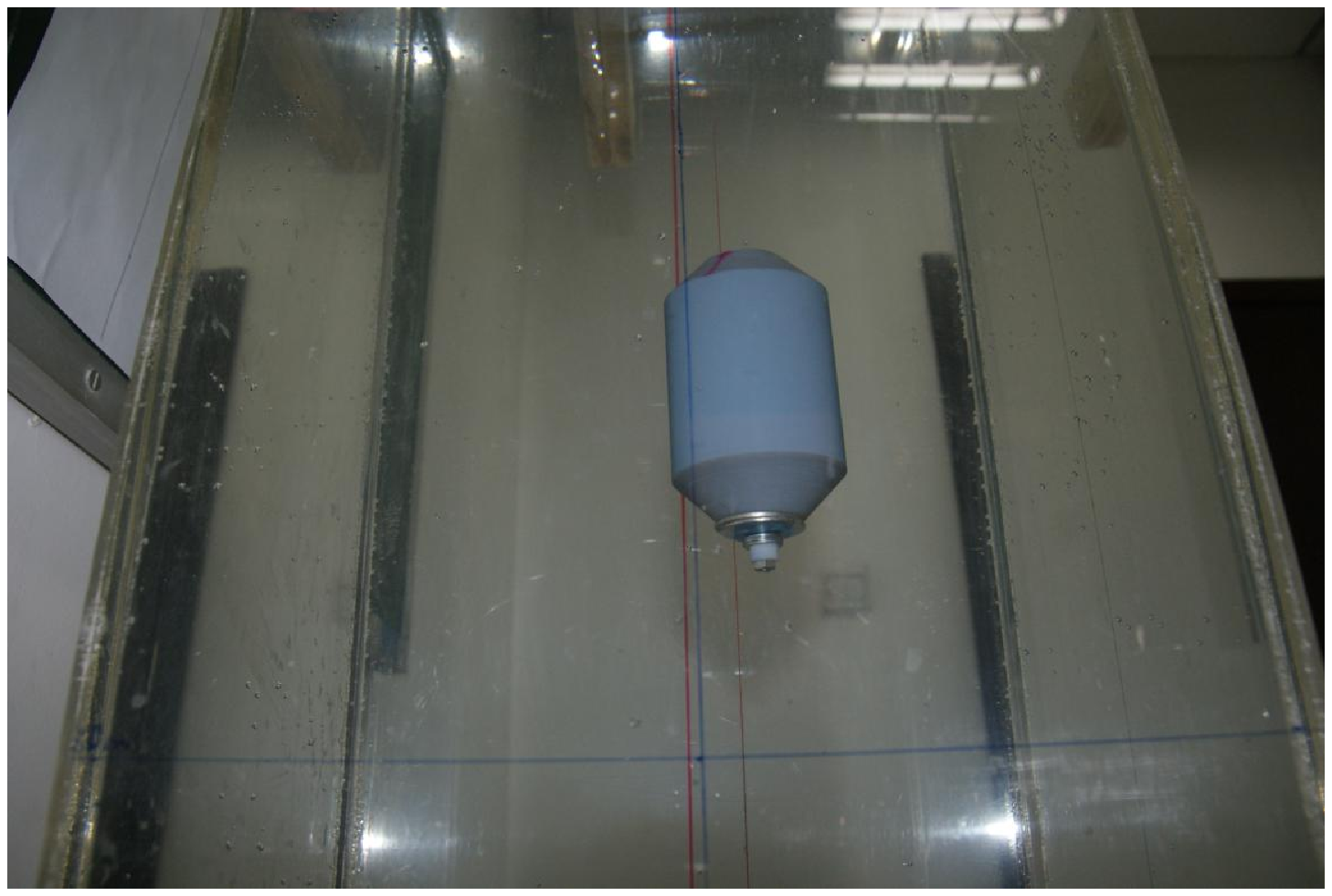}}
\vskip 3mm
\qleg{Fig. 2a: Ultra-slow fall.}
{The fall occurs in a water tank over a height of 1 meter 
and has a duration
of about 40 seconds. The diameter of the falling module is
61 millimeters, its height is 100mm
and its mass is 255 grams.
The water tank has a square section of $ 30\hbox{cm}\times 30\hbox{cm} $
and a height of 2 meters in two stages;
in this experiment only its first stage
was used. A fairly short fall duration was selected 
in order to minimize the effect of residual water currents.}
{}
\end{figure}
%

%
\begin{figure}[htb]
 \centerline{\psfig{width=7cm,figure=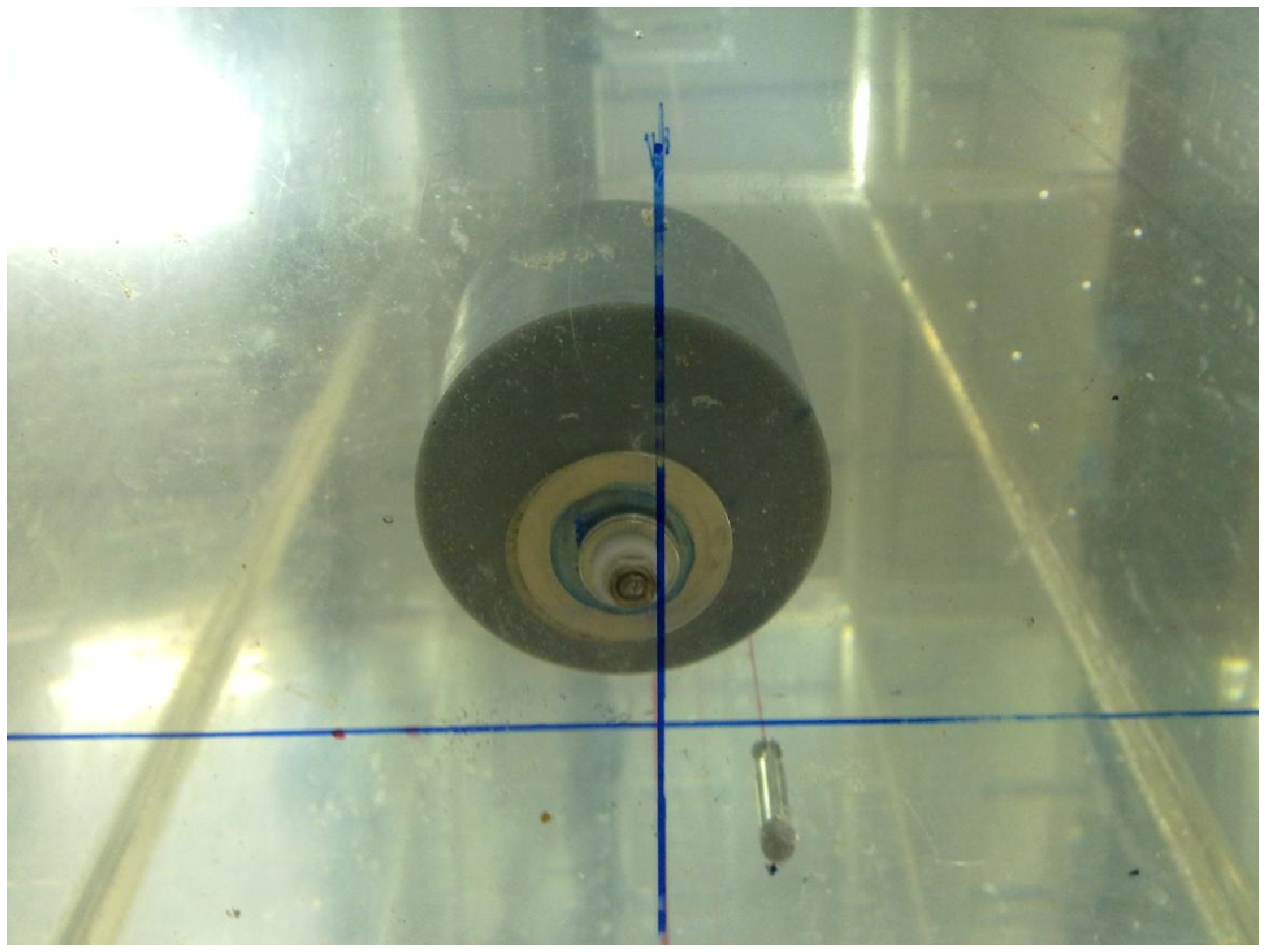}}
\vskip 3mm
\qleg{Fig. 2b: Picture of the module after it has reached the bottom
of the water tank.}
{At the bottom of the picture one can see a small plumb-bob
which is used to follow the horizontal deflections of the falling
body. In this experiment
the average eastward deviation over 40 falls was 
$ \Delta x_m = 2\pm 0.8 $mm; in this result the error bar
is the standard deviation of the average, that is to say
$ \sigma/\sqrt{40} $ where $ \sigma $ is the standard deviation
of the 40 measurements.
This result should be compared 
with the expected deviation which, in these conditions, is 
$ 1.8 $mm. 
In the air, the deviation would be about 100 times smaller.}
{}
\end{figure}
%
%
\qbu {\bf Start}\quad First, the module must be smoothly
released from a set position. In the air this a tricky problem
because even a low lateral initial velocity will develop into
a fairly large deflection during the duration of the fall.
In water this problem is much less serious because an initial
horizontal velocity will be reduced fairly quickly due to
friction. \qL
We have been using a system in which the start of the module
could be obtained by adding just one drop of water. It seemed
to work in a satisfactory way.
\qbu {\bf Residual currents}\quad When the module falls it creates
currents at different scales. Similarly currents
are created again when the module is
brought back to the surface for a new trial. How long
should one wait before starting the next fall? 30 s, 2 mn, 5 mn
or 10 mn. We do not know and 
this is a serious problem. \qL
As a matter of fact, there is
no single answer. If the module falls fairly quickly it is
much less affected by the currents than when it falls in two or three
minutes because (i) The currents do not have the time
to deflect it. (ii) When the vertical force experienced
by the module is fairly large it will not be much affected by the
small forces due to intermittent currents.
\qpar

With a waiting time of one minute, one observes
that the standard deviation of the landing points increases with the
duration of the fall%
\qfoot{For instance in one set of experiments the standard deviation
was 7\ mm for 65\ s and increased to 10\ mm for a duration of 100\ s.}%
.
As an extreme case a module which floats under water in a
neutral position will slowly drift from the middle of the tank to one
of its sides. \qL
A possible way to estimate the effect of the currents is to make several
series of measurements with identical falling times
but increasing waiting times between the falls.
\qbu {\bf Asymmetries of the module}\quad When the module is suspended
in water just before start, one can check whether or not
it is really vertical. Even objects which have been produced or
controlled on a lathe may not be completely vertical. When 
the axis of a module makes an angle $ \theta $ with the vertical,
it will have a tendency to follow that direction during its fall.
\qbu {\bf Turbulence}\quad  With a size of the order of 10cm and
a velocity of the order of 0.5cm/s, the Reynolds number
$ Re=\rho v D/\eta $ is of the order of 1,000 which means that
one is far from laminar flow conditions. Nevertheless, the 
dispersion which comes with this level of turbulence seems to
be fairly low with respect to other sources of noise. 
This can be seen from the observation that for
falls performed in succession 
under identical conditions the results change by ``batch'', 
by which we mean that after 5 or 6 falls having almost the same
arrival point, there may be a new batch of falls centered
around another landing point. This behavior is not yet clearly
understood but it does not seem to be due to turbulence.

\qA{Remedies}
The problem of residual currents can in principle
be overcome by waiting long enough between successive falls.
So, the most serious problem seems to be the asymmetries of the
module%
\qfoot{One solution may be to use a ``ball'' of oil in a solution
of ethanol. In this case the spherical shape
is assured by tension forces. However,
even in such a case there may be defects in the form of
small droplets of ethanol included into the ball of oil.}%
.
These asymmetries can be of a static or dynamic nature.
By static asymmetries we mean those which show up in 
the fact that the module is not completely vertical when 
floating without motion at the surface.
By dynamic asymmetries we mean those which result in 
asymmetrical friction forces which bring about 
spurious deflections during the fall of the module.
\qpar

One way to mitigate this difficulty is to control
for the angular orientation of the module at start and to
check that it does not rotate during the fall.
We followed this method when observing 40 falls in the sense
that there were equal numbers of falls done with 4 initial angular
orientations 90 degrees apart. 
In this way the effect of possible 
asymmetries should be averaged out. 
One drawback of this method 
is that it requires 4 times more trials.

\qA{Shape of the module}
What shape should one give to the falling module? There are two
conflicting requirements.
\qee{1} As noted previously the ratio $ V/A $ (where $ A $ is the
vertical section) must be large which suggests a fairly flat module.
\qee{2} However one must also minimize the deflections due to possible
asymmetries which would rather suggests a long module of small
horizontal section. 
\qpar
The challenge is to find a shape which fulfills these conflicting
requirements.

\vskip 4mm
{\bf Acknowledgment} The author expresses his gratitude to
Mr. Gen Li (Beijing Normal University) for providing
useful advice and for his help in making the pictures which
illustrate the article.

\appendix

\qI{Appendix A: Theoretical formula for eastward deflection}

For a body falling freely from a height $ h $ the 
eastward deviation is given by the formula 
(Cabannes 1966, p. 65 or Goldstein et al. 2004, p. 179)
 $$ d={ \Omega_h \over 3 }\sqrt{ { (2h)^3 \over g }},\quad 
 \Omega_h=\Omega \cos\lambda \qn{A1} $$

where:
\qbu 
$ \Omega $: vector of angular rotation of the Earth: 
$ \Omega = { 2\pi \over 3600\times 24 }=7.3\ 10^{-5} $ radian/second.
\qbu $ g $: acceleration of gravity, $ g=9.81 \hbox{ms}^{-2} $
\qbu $ h $: height of fall
\qbu $ \lambda $: latitude of the fall.
\qpar

This formula is not an exact result in the sense that it 
assumes that $ \Omega t $ is small with respect to 1
(Cabannes 1966, p. 65).
This means
that the duration of the fall should not
exceed $ 1/k\Omega $ ; with $ k $ of the order of 10, one
gets a limit of the order of half an hour.\qL
The second-order term in the development 
corresponds to a southward deflection whose order of magnitude is
$ \Omega t $ times the first-order term (Wikipedia 2011b).
This means that for times of the order of one minute
this southward deflection is some 1,000 times smaller than the
eastward deflection. In other words, this effect is much smaller
than the eastward deflection due to the rotation around the Moon
which is only 27 times smaller than the first-order effect. 
\qpar

Moreover it should be noted that 
that formula (A1) does not take into
account the friction of the falling body in the surrounding fluid.
This has two consequences, 
a first one which is
of little importance and a second which is much more serious.
\qee{1} The first consequence is that 
the friction will reduce the velocity. 
If one assumes that the velocity remains small with respect
to the limiting velocity this will result only in a slight change.
\qee{2}
Much more serious is the fact that friction creates turbulence
in the wake of the falling body. This will provoke random 
changes in the pressure-field around 
which in turn will provoke small random horizontal accelerations.
Because of its randomness turbulence cannot be modeled
in an appropriate way, which means that this effect cannot
be included in formula (A1). It may be useful to keep in mind
the following facts which result from empirical observation.
(i) Turbulence increases with velocity (ii) For a given velocity,
the effect of turbulence on the trajectory of the falling
body will be more serious when its mass becomes smaller. 

\qI{Appendix B: Experimental results in air}

Table B1 gives a summary of some of the historical experiments done
in air.

\begin{table}[htb]

\centerline{\bf Table B1\quad Deflection toward the east 
in free fall experiments}

\vskip 5mm
\hrule
\vskip 0.5mm
\hrule
\vskip 2mm

$$ \matrix{
\hbox{Physicist} \hfill & \hbox{Year}  & \hbox{Location}  \hfill
& \hbox{Height}  
& \hbox{Eastern} & \hbox{Accuracy} & \hbox{Predicted}\cr
\hbox{} & \hbox{}  & \hbox{}  & \hbox{}  
& \hbox{deflection} & \hbox{}  & \hbox{deflection}\cr
\hbox{} & \hbox{}  & \hbox{}  & \hbox{}  
& \hbox{} & \hbox{}  & \hbox{(in vacuum)}\cr
\qtb 
\hbox{} & \hbox{}  & \hbox{}  & \hbox{[m]}  
& \hbox{[mm]} & \hbox{[\%]} & \hbox{[mm]}\cr
\noalign{\hrule}
\qth
\hbox{Guglielmini} \hfill& 1791  & \hbox{Bologna (44.50)} \hfill
 & \hfill 78  & 18.9 &  & 10.7\cr
\hbox{Benzenberg}\hfill & 1802  & \hbox{Hamburg (53.57)} \hfill 
&\hfill 76  & 9.0 & & 8.61 \cr
\hbox{Reich} \hfill&  1831 & \hbox{Freyberg (48.20)} \hfill & 
\hfill 158  & 28.4\pm 0.4 & 1.4\%
& 29.0\cr
\hbox{Hall}\hfill & 1903  & \hbox{Boston (42.31)}\hfill& \hfill 23& 1.49\pm
0.05 & 3.3\% & 1.78 \cr
\qtb
\hbox{B\"ahr et al.} \hfill& 2005  & \hbox{Bremen (53.06)}\hfill
& \hfill 145 & 26.4\pm 1.4 & 5.3\%
& 23.0 \cr
\noalign{\hrule}
} $$
\vskip 1mm
Notes: Latitudes (expressed in degrees) are given
within parenthesis in the location column. 
The number of falls performed were as follows: Benzenberg:
32, Reich: 106, Hall: 946, B\"ahr et al.: 120.
Formula (A1) for a fall in vacuum which was used to compute
the numbers in the last column does not include the effect of air
friction. Therefore
the accuracy column does not refer to accuracy with
respect to this formula; it gives the ratio of the measurement
error (i.e. the dispersion of the impacts) to the average
deflection.\qL
The probable error indicated as $ \pm $ is given in Bruhat (1955)
for the Reich experiment and in Hall (1903b) for the Hall
experiment. Actually, Hall gives two  
probable errors (depending on differing assumptions): 0.05mm 
and 0.15mm.\qL 
Identification of the Coriolis force due to the rotation
of the Earth around the center of gravity of the Earth-Moon system
requires an accuracy better than $ 1/21=4\% $. This would well 
have been 
within the reach of the Reich experiment. However, as explained 
in the text, the error given for the Reich is probably 10 times
too small.
\qL
{\it Sources: Benzenberg (1804), Bruhat (1955), Hall (1903 a,b)} 
\end{table}

\qI{Appendix C: Effect of turbulence in air}

In previous papers (Poujade 2010, Roehner 2010)
it was shown that for Reynolds numbers
over 1,000 the radius of dispersion for deflections 
of spheres due to
turbulence can be represented by the following semi-empirical
formula:
$$ \Delta = 0.02{ \rho \over \rho_1 }{ H^{3/2} \over \sqrt{r} } \qn{C1} $$

where:\qL
$ \Delta $: Dispersion radius due to turbulence\qL
$ \rho $: Density of the fluid in which the spheres are falling\qL
$ \rho_1 $: Density of the spheres\qL
$ H $: Height from which the spheres are falling\qL
$ r $: radius of the spheres
\qpar

Formula C1 was tested for falls of several meters 
in air and was found fairly satisfactory. However, it may
not be applicable to 
the conditions of ultra-slow falls in a liquid%
\qfoot{In such conditions one has $ \rho \simeq \rho_1 $.
If in addition 
one takes for instance $ r=10 $cm and $ H=1 $m, one
gets $ \Delta\sim 10 $cm, which seems much too large.}%
.
\qpar

It can be observed that both the eastward deviation (given in (1b))
and
the dispersion due to turbulence contain the factor $ H^{3/2} $
which means that one does not gain much in terms of
accuracy by increasing $ H $. What makes small values of $ H $ 
inconvenient and unpractical is rather the limitation of
accuracy in the determination of the vertical. It is probably 
difficult to determine the vertical
with an accuracy better than 1/100 of a millimeter. Thus, if
one wishes a precision better than 1\% the deflection must 
be at least 1mm which means a height of more than 10 meters.
\qpar

If one applies formula (C1) to the Reich and
Hall experiments one gets: $ \Delta= 44 $mm and $ \Delta=2.5 $mm
respectively.
The accuracy on the center (defined as the average of the
coordinates of individual impacts) then depends on the
number of balls that have been used. If one can assume that
successive balls fall independently 
\qfoot{Which means that the perturbation due to the fall of ball
number $ n $ has been sufficiently dampened so as not
to  affect the fall of ball number $ n+1 $.}
the standard deviation on the average of $ N $ balls will be
$ \epsilon=\Delta /\sqrt{N} $.
\qpar

For the Hall experiment this gives: 
$ \epsilon = 0.08/\sqrt{946}=0.08 $mm
which is comprised between the two error levels given
by Hall, namely  0.05mm and 0.15mm.\qL
For the Reich experiment one gets: $ \epsilon=44/\sqrt{106}=4.4 $mm.
This is about 10 times more than the probable error given for
this experiment and raises some doubts about the claimed accuracy.
\qpar

The dispersion radius given by (C1) corresponds to random
deflections which can, at least in principle, be eliminated
by taking an average over a sufficiently large sample of falls.
This averaging process can be illustrated by the Hall 
experiment. With $ \epsilon =0.08 $mm, the accuracy 
of the measurement is $ 0.08/1.5=5.4\% $. If one would like to
increase the accuracy to 1\% one would have to perform $ 5.4^2=29 $
times more falls. As Hall already performed some 1,000 falls,
this means that one would have to perform 29,000 falls.
This is almost an impossible task except perhaps if it can
be made completely automatic. However, it should be noted that by
using platinum balls ($ \rho_1=21.5 $kg/liter) instead of bell-metal
($ rho_1=8 $kg/liter) one would gain a factor $ 21.5/8=2.7 $.
that would reduce the number of falls to 10,700; still a 
large number.

\vskip 10mm

{\bf \large References}
\vskip 5mm

\qparr
Benzenberg (J.F.) 1804: Versuche \"uber das Gesetz des Falls, \"uber
den Widerstand der Luft und \"uber die Umdrehung der Erde.
Nebst des Geschichte aller Fr\"uheren Versuche von Galil\"ai bis auf
Guglielmini.

\qparr
B\"ahr (J.), Log\'e (M.),
Mechelk (K.), Operhalsky (A.-M). 2005:  ``Dreht sich die Erde?''
[Account of an experiment made at a German space research 
installation allowing free fall in vacuum
from a height of 145 meters; available on the Internet].

\qparr
Bruhat (G.) 1955: M\'ecanique. Masson, Paris.

\qparr
Cabannes (H.) 1962, 1966: Cours de m\'ecanique g\'en\'erale.
Dunod, Paris.

\qparr
Goldstein (H.), Poole (C.), Safko (J.) 2002: Clasical
mechanics. 3rd edition. Addison-Wesley.

\qparr
Hall (E.H.) 1903a: Do falling bodies move south?
Part I: Historical.
Physical Review Vol. 17, no 3 (September) p. 179-190.\qL
[Surprisingly a
this historical review article gives almost no references to 
previous publication. The only reference that is given is
to an article published in 1802 by Benzenberg in
Annalen der Physik (Vol. 12, p. 367-373).]

\qparr
Hall (E.H.) 1903b: Do falling bodies move south?
Part II: Methods and results
Physical Review Vol. 17, no 4, (October) p. 245-254.

\qparr
Poujade (O.) 2010: Impact location broadening of a solid
object falling in a fluid due to drag and lift fluctuations.
CEA (22 January 2010).\qL
Unpublished paper, private communication. 

\qparr
Roehner (B.M.) 2010: Exploring turbulence behind a sphere
through free-fall experiments. Part I: Spheres falling in air.
Working report LPTHE (9 May 2010).

\qparr
Wikipedia 2011a:  Article entitled ``Terminal velocity''.

\qparr
Wikipedia 2011b: French version of the article entitled
``D\'eviation vers l'est'' (i.e. Eastward deflection).

\end{document}

%% file: cpreb.tex
\usepackage{epsfig,times,lscape}
\usepackage[usenames]{color}

    \ifnum\count102=1

    \fi

    \ifnum\count102=2
\fi

\newcommand{\nc}{\newcommand}

     \ifnum\count101=1
\nc{\qI}[1]{\section{{#1}}}
\nc{\qA}[1]{\subsection{{#1}}}
\nc{\qun}[1]{\subsubsection{{#1}}}
\nc{\qa}[1]{\paragraph{{#1}}}

\def\qpar{\vskip 2mm plus 0.2mm minus 0.2mm}
\def\qL{\hfill \break}
     \fi 

      \ifnum\count101=2
 \nc{\qI}[1]{\parindent=0mm \vskip 8mm 
{\centerline{\LARGE \color{red}#1}}\vskip 3mm}
\nc{\qA}[1]{\vskip 2.5mm \noindent {{\bf        #1}} \vskip 1mm
\parindent=0mm}
 \nc{\qun}[1]{\vskip 1mm \noindent {\sl #1 }\quad }

\def\qL{\hfill \break}
\def\qpar{\vskip 2mm plus 0.2mm minus 0.2mm}

      \fi

\def\qth{\vrule height 12pt depth 0pt width 0pt}
\def\qtb{\vrule height 0pt depth 5pt width 0pt}

\nc{\qfoot}[1]{\footnote{{#1}}}

      \ifnum\count101=1
\def\qbu{\hfill \par \hskip 6mm $ \bullet $ \hskip 2mm}
\def\qee#1{\hfill \par \hskip 6mm (#1) \hskip 2 mm}
      \fi
      \ifnum\count101=2
\def\qbu{\hfill \par \hskip 4mm $ \bullet $ \hskip 2mm}
\def\qee#1{\hfill \par \hskip 4mm (#1) \hskip 2 mm}
      \fi

\def\qparr{ \vskip 1.0mm plus 0.2mm minus 0.2mm \hangindent=10mm
\hangafter=1}

     \ifnum\count101=1 
 
     \fi
     \ifnum\count101=2

  \def\qcitb#1{\noindent \hbox to 102mm{\hfill \small #1} \vskip 1mm}
      \fi

 \def\qpages#1{\count102=0{\loop\advance\count102 by 1
 \null \vfill\eject \ifnum\count102<#1 \repeat}}

\def\qn#1{\eqno \hbox{(#1)}}

\def\qth{\vrule height 12pt depth 0pt width 0pt}
\def\qtb{\vrule height 0pt depth 5pt width 0pt}

\def\qv{\vskip 0.1mm plus 0.05mm minus 0.05mm}

\def\qhw{\hskip 1.5mm}
\def\qleg#1#2#3{\noindent {\bf \small #1\qhw}{\small #2\qhw}{\it \small #3}\qv }
